\documentclass[a4paper]{jpconf}
\usepackage{graphicx,bm}
\begin{document}
\title{
Longitudinal magnetic excitation in KCuCl$_{3}$ 
studied by Raman scattering under hydrostatic pressures
}

\author{H Kuroe$^{1}$, N Takami$^{1}$, N Niwa$^{1}$, T Sekine$^{1}$, 
M Matsumoto$^{2}$,\\ F Yamada$^{3}$, H Tanaka$^{3}$ and K Takemura$^{4}$}

\address{%
$^{1}$Department of Physics, Sophia University, Tokyo 102-8554, Japan%
}
\address{%
$^{2}$Department of Physics, Shizuoka University, Shizuoka 422-8529, Japan%
}
\address{%
$^{3}$Department of Physics, Tokyo Institute of Technology, Tokyo 152-8551, Japan%
}
\address{%
$^{4}$National Institute for Materials Science (NIMS), Tsukuba, Ibaraki 305-0044, Japan%
}

\ead{kuroe@sophia.ac.jp}

\begin{abstract}
We measure Raman scattering in an interacting spin-dimer system 
KCuCl$_{3}$ under hydrostatic pressures up to 5 GPa mediated by He gas.
In the pressure-induced quantum phase, 
we observe a one-magnon Raman peak, 
which originates from the longitudinal magnetic excitation
and is observable through the second-order exchange interaction Raman process.
We report the pressure dependence 
of the frequency, halfwidth and 
Raman intensity of this mode.
\end{abstract}

\section{Introduction}

The magnetic-field and pressure-induced quantum phase transitions 
in TlCuCl$_{3}$ and KCuCl$_3$ have been extensively 
studied.\cite{Tanaka1996,Takatsu1997}
In the spin-gapped phase, 
the $S = 1/2$ interacting antiferromagnetic (AF) spin dimer system 
has a spin-triplet excited state 
separated from the spin-singlet ground state.
Because of the magnetic interaction between spin dimers, 
the excited state is dispersive and then 
the spin-gap energy is much lower than 
that of the isolated spin dimers.
This is described by the bond-operator model.\cite{Matsumoto2004}
The Zeeman effect under a magnetic field, 
the increase of interdimer interaction under high pressure 
and/or the impurity doping 
cause the quantum phase transition to the ordered phase.
The ordered phase is characterized by 
the absence of the spin gap and 
the AF ordered moment of which amplitude can fluctuate.
This fluctuation is the longitudinal magnetic excitation 
which we focus on in this paper.

As well as inelastic neutron scattering (INS), 
the inelastic light scattering, 
known as Raman scattering, 
is a powerful tool to detect the
longitudinal magnetic excitation.\cite{Matsumoto2008}
In the quantum phase induced by magnetic field, 
the longitudinal magnon excitation is detected 
as a one-magnon Raman peak 
induced through the second-order exchange interaction 
Raman process.\cite{Kuroe2008, Kuroe2008ICORS}
Comparing INS, only a small size of sample 
is enough to measure Raman scattering.
Using the diamond anvil cell (DAC), 
we can apply hydrostatic pressure up to about 10 GPa.
These are advantages of Raman scattering.
In this paper, 
we report the one-magnon Raman scattering 
in the pressure-induced ordered phase of KCuCl$_3$.

\section{Experiments}
The single crystal of KCuCl$_{3}$ 
was grown from a melt by the standard Bridgeman technique.
A small piece of it 
with a size of 100 $\times$ 50 $\times$ 10 $\mu$m 
was enclosed into DAC.
And then, the DAC was placed in a continuous flow type cryostat 
to cool the single crystal down to 3 K.

We used He gas as a pressure medium 
because it gives almost hydrostatic conditions.\cite{Takemura2008,footnote}%
Moreover, the single crystal is easily dissolved 
in methanol-ethanol mixture, 
which is a standard pressure medium.
The pressure $P$ was monitored 
by using the wavelength shift of 
the $R_{1}$ fluorescence line of a ruby powder.
Because the wavelength of the $R_{1}$ fluorescence line 
depends on temperature, 
we set ruby powder in high pressure section 
and in the ambient pressure one (on the back surface of diamond).
Because we can detect the $R_{1}$ fluorescence peaks 
from the ruby powder at the high pressure section 
and that at the ambient pressure one simultaneously, 
this method estimates very precise pressure at low temperatures.

Raman scattering was excited by an Ar$^{+}$-ion laser 
with the wavelength of 514.5 nm.
The scattered light was dispersed 
by using a triple grating monochromator (Jobin-Yvon T64000)
and was detected by a CCD detector cooled by liquid-N$_{2}$.
In order to observe the magnon Raman spectra strongly, 
the incident laser was polarized 
along the direction of the strong interdimer interaction, 
which is almost parallel to the one 
along the intradimer interaction,\cite{Kuroe2008}
and we did not use the analyzer (polarizer) 
in front of the entrance slit of the monochromator.
The Raman shift was calibrated 
by using a rotational Raman spectrum of air.

\section{Results} 
\begin{figure}
\includegraphics[width=\textwidth]{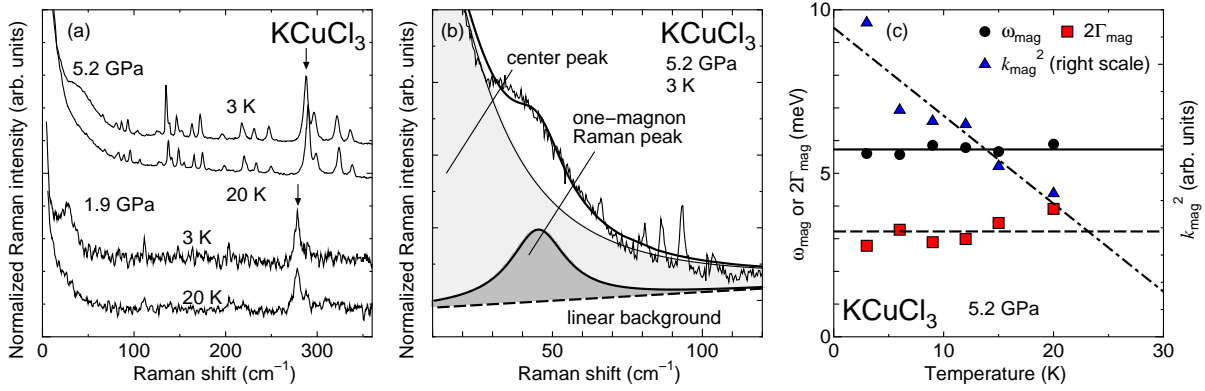}
\caption{
(a) Raman spectra at 3 and 20 K under pressures of 1.9 and 5.2 GPa.
(b) Low-energy Raman spectrum at 3 K under 5.2 GPa 
and its decomposition.
(c) Temperature dependences of the frequency $\omega_{\rm mag}$, 
halfwidth $2\Gamma_{\rm mag}$ and squared coupling coefficient 
$k^2_{\rm mag}$ of the one-magnon Raman peak at 1.9 GPa.
The details of the fitting are written in text.
}
\end{figure}
Figure 1(a) shows Raman spectra at 3 and 20 K 
under pressures of 1.9 and 5.2 GPa.
The Raman intensity was normalized 
by the integrated intensity 
of the Raman peak around 300 cm$^{-1}$ in each spectrum, 
which is indicated by an arrow in Fig. 1(a).
The overall phonon Raman spectrum above 50 cm$^{-1}$ is consistent 
with the one in the previous report at ambient pressure.\cite{Choi2005}
The frequencies of the phonon peaks increase 
with increasing pressure.
Using this fact, we can obtain the mode Gr\"{u}neisen constant.
However, it is beyond of the scope 
of this paper and will be presented elsewhere.
One can see that the Raman peak around 25 cm$^{-1}$ at 3 K under $P$ = 1.9 GPa 
is superimposed on the strong tail centered at 0 cm$^{-1}$
which comes from the stray light of the reflection 
and the Rayleigh scattering.
This peak was not observed at 20 K.
As will be discussed in the following, 
this is the one-magnon Raman peak 
from the longitudinal magnon excitation.

Figure 1(b) describes the fitting procedure at $P$ = 5.2 GPa.
For quantitative discussion, 
we extract the one-magnon Raman peak $I_{\rm mag}(\omega)$ 
from the observed Raman spectrum $I_{\rm obs}(\omega)$, 
by using the nonlinear least square fit with the model
of two Lorentz functions on a linear background:
\begin{equation}
I_{\rm obs}(\omega)
\equiv I_{\rm mag}(\omega) 
+ \frac{k_{\rm c}^{2} \Gamma_{\rm c}}{\omega^{2}+\Gamma_{\rm c}^{2}}
+ (a \omega + b)
= \frac{[n(\omega_{\rm mag}) + 1]k_{\rm mag}^{2} \Gamma_{\rm mag}}
       {(\omega-\omega_{\rm mag})^{2}+\Gamma_{\rm mag}^{2}}
+ \frac{k_{\rm c}^{2} \Gamma_{\rm c}}{\omega^{2}+\Gamma_{\rm c}^{2}}
+ (a \omega + b)
\ , 
\end{equation}
where $\omega_{\rm mag}$, $k_{\rm mag}$ ($k_{\rm c}$), 
$\Gamma_{\rm mag}$ ($\Gamma_{\rm c}$), $n(\omega_{\rm mag})$ 
and $(a \omega + b)$ denote 
the frequency, the coupling coefficient and the halfwidth  
of the one-magnon mode (the strong tail at 0 cm$^{-1}$), 
the Bose factor and the linear background, respectively, 

Figure 1(c) shows the temperature dependences of 
$\omega_{\rm mag}$, $2\Gamma_{\rm mag}$ and 
the squared coupling coefficient 
$k_{\rm mag}^{2}$.
One can see that $k_{\rm mag}^{2}$ decreases 
with increasing temperature, 
while $\omega_{\rm mag}$ and $2\Gamma_{\rm mag}$ are almost 
independent of temperature.
And therefore, 
we estimate $k_{\rm mag}^{2}$, $\omega_{\rm mag}$ and $2\Gamma_{\rm mag}$ 
at zero temperature.
With the values obtained under different pressures, 
we show the pressure dependences of 
$\omega_{\rm mag}$, $2\Gamma_{\rm mag}$ and $k_{\rm mag}^{2}$ 
as functions of applied pressure in Fig. 2.

\section{Discussion}
\begin{figure}
\begin{minipage}{0.55\textwidth}
\begin{center}
\includegraphics[width=0.7\textwidth]{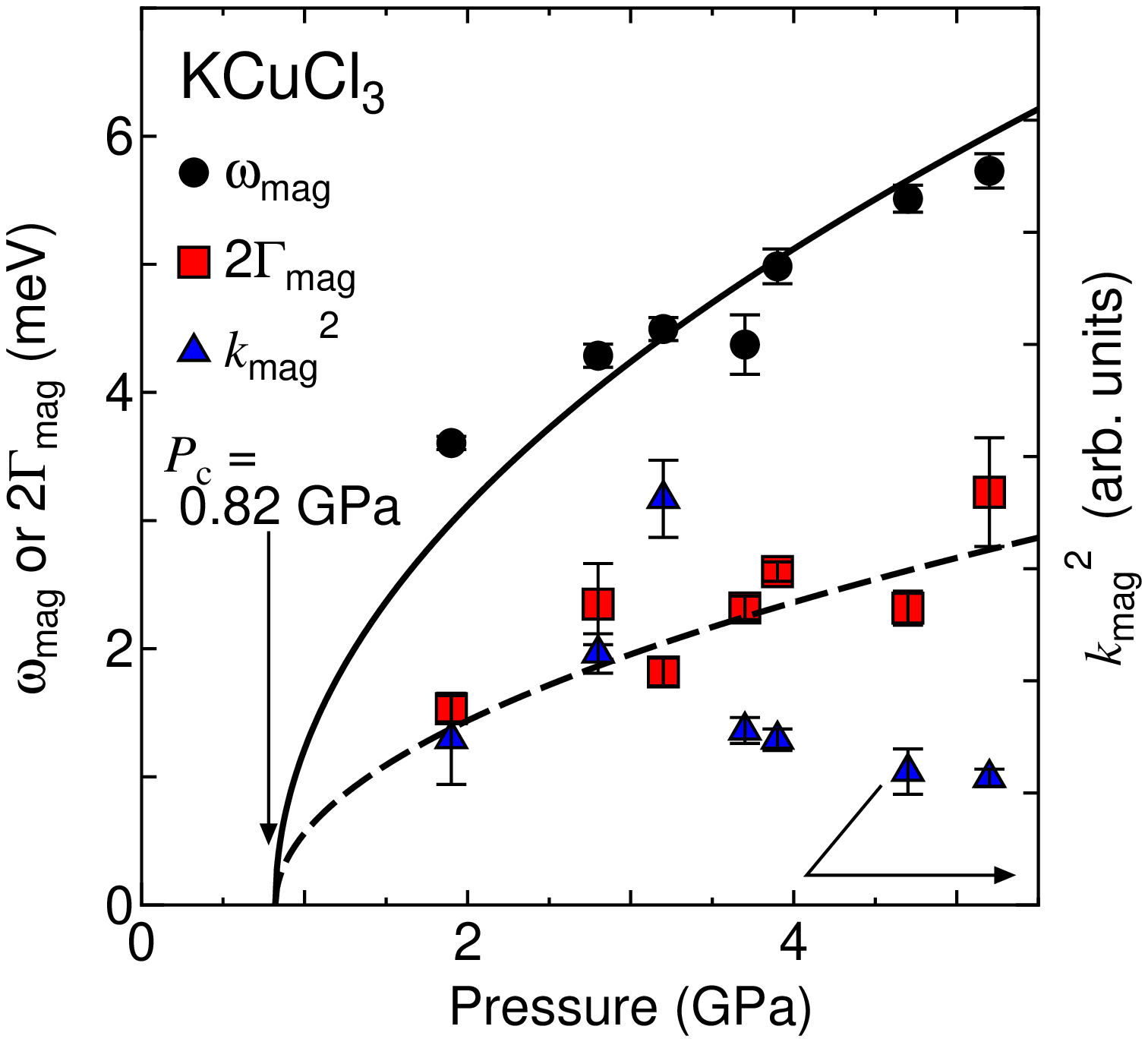}
\caption{
Pressure dependences of the frequency $\omega_{\rm mag}$, 
halfwidth $2\Gamma_{\rm mag}$ and squared coupling coefficient 
$k^2_{\rm mag}$
of the one-magnon Raman peak.}
\end{center}
\end{minipage}
\hspace*{0.05\textwidth}
\begin{minipage}{0.4\textwidth}
\begin{center}
\includegraphics[width=0.9\textwidth]{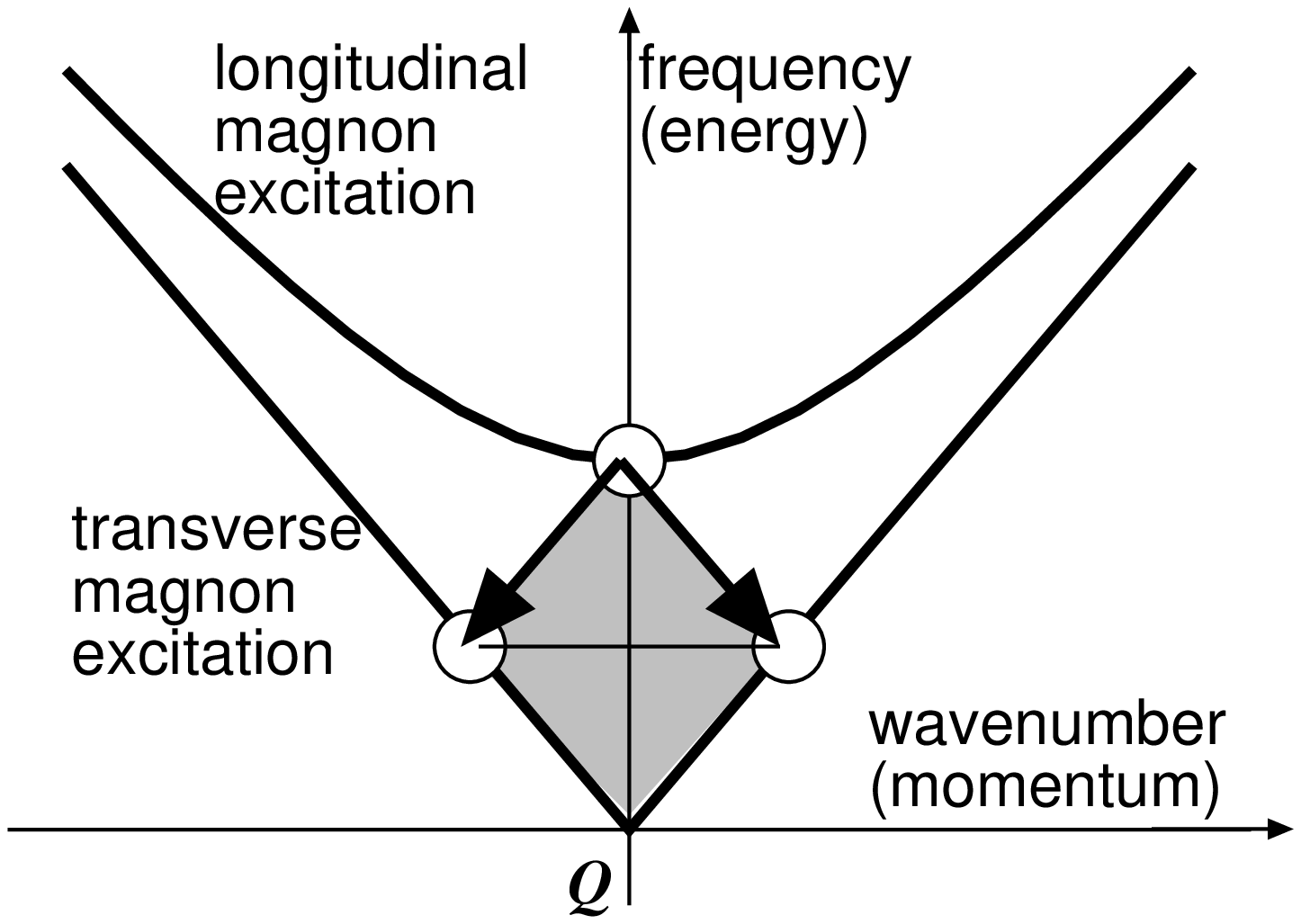}
\caption{
Schematic of a decay process from 
one longitudinal magnon to 
two transverse magnons 
in the energy-momentum space. 
${\bm Q}$ denotes the magnetic zone center.
}
\end{center}
\end{minipage}
\end{figure}

We observed that
the frequency and halfwidth of the one-magnon Raman peak are 
proportional to a function of $(1-P/P_{\rm c}^{\rm K})^{1/2}$, 
where the critical pressure 
$P_{\rm c}^{\rm K}$ (= 0.82 GPa) was obtained 
by magnetization measurement 
under hydrostatic pressures.\cite{Goto2006} 
First, we compare these results to 
the results of INS in TlCuCl$_{3}$ 
taken under hydrostatic pressures.\cite{Ruegg}
The frequency and halfwidth of the longitudinal magnetic excitation
are proportional to a function of $(1-P/P_{\rm c}^{\rm Tl})^{1/2}$ in TlCuCl$_{3}$, 
where $P_{\rm c}^{\rm Tl}$ $\sim$  0.1 GPa.
Then we conclude that the one-magnon Raman peak 
observed at low temperatures under higher pressures than 0.82 GPa 
comes from the longitudinal 
magnetic excitation 
through the second-order exchange interaction 
magnon Raman process.\cite{Matsumoto2008}

We focus on the halfwidth $\Gamma_{\rm mag}$ 
proportional to the frequency $\omega_{\rm mag}$.
Because $2\Gamma_{\rm mag}$ corresponds to the inverse lifetime of magnon, 
we consider the decay process of the longitudinal magnetic excitation.
In the bond-operator model, 
the terms of two spin scalar product 
in the spin Hamiltonian and the effective magnon Raman operator are expressed 
as the quadratic terms of 
the singlet and triplet operators.\cite{Matsumoto2008,Kuroe2008,Kuroe2008ICORS}
In the ordered phase, 
these operators are mixed with each other and 
the system can be rewritten by using 
the quadratic terms of the four mixed singlet-triplet operators.
Because one of these four can be treated as 
the uniformly condensed mean-field operator 
($c$-number operator, $\overline{a}$ in ref. \cite{Matsumoto2004}), 
the interdimer interaction can be written 
by using three kinds of operators 
($b_{{\bm k}{\rm L}}$, $b_{{\bm k}y}$ and $b_{{\bm k}z}$ in ref. \cite{Matsumoto2004}, 
where $\hbar{\bm k}$ is the momentum of magnon).
Here $b_{{\bm k}{\rm L}}$ is 
the annihilation operator of the longitudinal magnetic excitation and 
the other two are those of the transverse ones.\cite{Matsumoto2004,Matsumoto2008}
The spin Hamiltonian and the effective Raman operator contain the terms of 
${\cal O}(b)$, ${\cal O}(b^2)$, ${\cal O}(b^3)$ and ${\cal O}(b^4)$, 
where $b$ denotes one of $b_{{\bm k}{\rm L}}$, $b_{{\bm k}{\rm L}}^{\dagger}$, 
$b_{{\bm k}y}$, $b_{{\bm k}y}^{\dagger}$, $b_{{\bm k}z}$ and $b_{{\bm k}z}^{\dagger}$.
To minimize the magnetic energy, 
the ${\cal O}(b)$ term in the spin Hamiltonian should vanish.
We note here that the ${\cal O}(b)$ term in the Raman operator can be finite
even in this case.
This is the origin of the one-magnon Raman 
scattering.\cite{Matsumoto2008,Kuroe2008}
The magnon dispersion curve comes from the ${\cal O}(b^{2})$ terms 
in the exchange interaction.\cite{Matsumoto2004}
From an analogy with the anharmonicity problem of phonons, 
one can easily show that 
the ${\cal O}(b^{3})$ term 
gives the finite lifetime of magnon 
due to the  magnon-magnon decay in Fig. 3, 
which is omitted in refs. \cite{Matsumoto2004} and \cite{Matsumoto2008}.
 
We consider the magnon-magnon decay channel 
discussed by Kulik and Sushkov \cite{Kulik} 
for the lifetime of the longitudinal magnetic excitation 
at the magnetic $\Gamma$ point ${\bm Q}$.
This Raman-active mode is massive and 
its dispersion curve around ${\bm Q}$
is proportional to $|{\bm Q}|^{2}$
with the minimum energy at ${\bm Q}$.
Below this energy, 
there exists only two branches of the magnetic excitations, 
i.e., two transverse modes.
These are the Goldstone modes 
of which dispersion curve 
around ${\bm Q}$ have a linear function of $|{\bm Q}|$.
The possible magnon-magnon decay channel 
can be found easily by using the geometric procedure 
on the energy-momentum space, as shown in Fig. 3.
Because the number of decay channels 
is proportional to $\omega_{\rm mag}^2$, 
the transition probability for one channel should be 
proportional to $\omega_{\rm mag}^{-1}$, 
resulting in the halfwidth 
proportional to $\omega_{\rm mag}$.
These can be calculated by using the bond-operator method, 
which will be published elsewhere.

The bond operator theory predicts that 
$k_{\rm mag}^{2} 
\propto (1-P/P_{\rm c}^{\rm K})^{1/2}$,\cite{Matsumoto2008}
which is inconsistent with our observation.
The strong tail at 0 cm$^{-1}$ 
might be a possible origin of this deviation.
As we stated, we normalized the squared coupling coefficient 
by that of the phonon mode around 300 cm$^{-1}$, 
i.e., we implicitly suppose 
the pressure independent properties of this phonon.
We omit the Raman intensity 
through the three magnon Raman process, 
which was observed at ambient pressure.\cite{Choi2005}
These are other possible origins
of this deviation.


\section*{Acknowledgement}
This work was partly supported 
by Grants-in-Aid for Scientific Research (C) 
(Nos. 22540350, 21550029 and 23540390) 
from the Ministry of Education, Culture, Sports, 
Science and Technology of Japan (MEXT).


\begin{thebibliography}{99}
\bibitem{footnote}
More precisely, 
we need to use the term of quasi-hydrostatic pressure.
However, because of the strong zero-point oscillation, 
high pressure He gas serves very high hydrostaticity 
even when it is frozen at low temperature.
\end{thebibliography}
\end{document}